\documentclass[aps,prl,reprint,nofootinbib]{revtex4-1}

\usepackage{hyperref} 

\usepackage{epsfig}
\usepackage{bm}
\usepackage{amssymb}
\usepackage{mathrsfs}
\usepackage{amsmath}
\usepackage{dsfont}
\usepackage{color}
\usepackage{cancel}
\usepackage{booktabs}
\usepackage{upgreek}

\newcommand{\be}{\begin{equation}}
\newcommand{\ee}{\end{equation}}

\newcommand{\C}{\mathcal C}

\newcommand{\Jxy}{J_{\text{xy}}}
\newcommand{\JI}{J_{\text{I}}}
\newcommand{\Latt}{\Uplambda}

\begin{document}
\title{Ising model as a \texorpdfstring{$U(1)$}{} Lattice Gauge Theory with a \texorpdfstring{$\theta$}{}-term}
\author{Tin Sulejmanpasic}
\email{tin.sulejmanpasic@durham.ac.uk}

\affiliation{Department of Mathematical Sciences, Stockton Road, Durham University, DH1 3LE Durham, United Kingdom\\
}

\begin{abstract}
We discuss a gauged XY model a $\theta$-term on an arbitrary lattice in 1+1 dimensions, and show that the theory reduces exactly to the 2d Ising model on the dual lattice in the limit of the strong gauge coupling, provided that the topological term is defined via the Villain action. We discuss the phase diagram by comparing the strong and weak gauge coupling limits, and perform Monte Carlo simulations at intermediate couplings. We generalize the duality to higher-dimensional  Ising models using higher-form U(1) gauge field analogues. 
\end{abstract}

\maketitle

The Ising model is one of the most important statistical mechanical models. Its simplicity, universality and exact solvability in 1d and 2d are just some of the reasons in its pervasiveness in physics. On the other hand, gauge theory is best known as the theory of light, which can be described by a gauge field space-time vector field $A_\mu$. We will be concerned here with $U(1)$ gauge theories. The fundamental principle underlying $U(1)$ gauge theories is gauge invariance, i.e. the statement that all observables are invariant under the gauge transformation $A_\mu\rightarrow A_\mu+\partial_\mu\varphi$, where $\varphi$ is an arbitrary angle-valued function of space-time coordinates (i.e. the space of gauge transformations is a circle, which is the same as the $U(1)$ group, hence the name). Crucially the k-charge Wilson loops $e^{ik\oint _{C}dx^\mu A_\mu}$, with a closed spacetime contour $C$ are invariant under gauge transformations only if the charge $k$ is an integer.  The space-time contour $C$ has an interpretation of a probe particle worldline, carrying $k$ units of $U(1)$ charge. 

\begin{figure}[htbp] 
   \centering
   \includegraphics[width=2in]{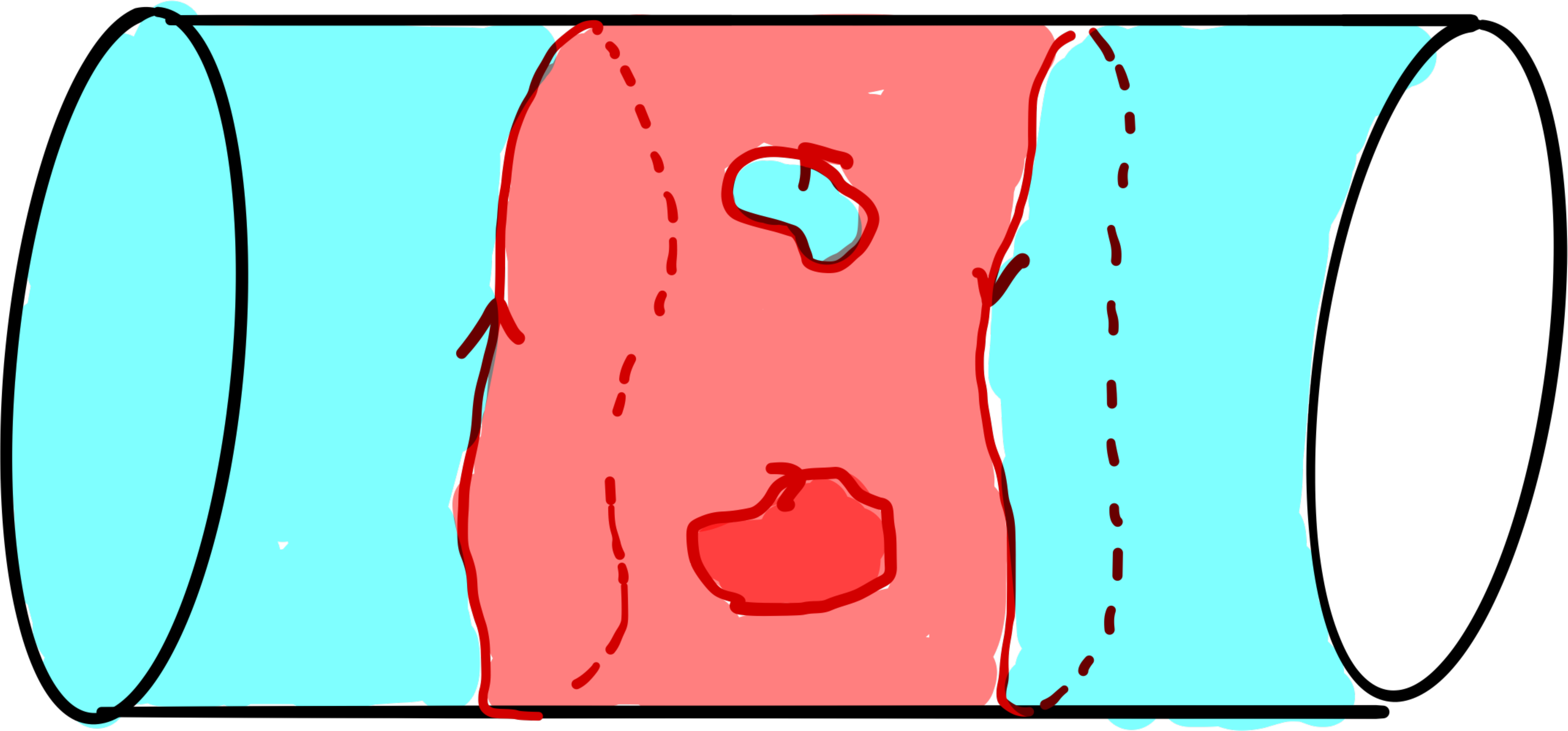} 
   \caption{A cartoon of the 1+1 gauge theory ensemble. The circle is the compact Euclidean time, and spatial direction extends from left to right. The blue areas are the ``vacuum'', while the red areas are vacuum excitations (i.e. strings) costing finite energy per area due to the electric field between the sources. If the external electric field is introduced (i.e. the $\theta$-term, the vacuum energy goes up, while the string tension, for appropriately oriented electric dipoles, goes down. At $\theta=\pi$ they become degenerate, and the elementary charges are no longer confined.}
   \label{fig:loops}
\end{figure}

In (1+1)d the connection of the gauge theory with the Ising model can be made evident by considering a single scalar field coupled to the $U(1)$ gauge field and was noted before (see e.g.~\cite{Affleck:1991tj,Komargodski:2017dmc}). The phenomenology of this model was discussed by Coleman before still \cite{Coleman:1978ae}. Firstly there exist two, potentially different, regimes\footnote{The two regimes are in fact not separated by a phase transition, unless the $\theta$-angle is set to $\pi$.}: the confining regime and the Higgs regime (see also \cite{Witten:1978bc}). In the deep confining regime, the mass-squared $M^2$ of the scalars is positive and large, and the theory is very nearly a pure gauge theory. Such a theory has massive excitations of $\phi$, which can be related to worldlines coupling to the gauge field via the Wilson loop $e^{i\oint dx^\mu A_\mu}$, wrapping in the Euclidean compact time direction. However the gauge field fluctuations impose a confining potential on these excitations, and the excitations have to pay the energy price of the string attached to the wordline. In addition the ensemble can be thought of as consisting of tiny loops of scalar matter, which renormalize the string tension (see Fig.~\ref{fig:loops}). 

One can then think about placing the system in the external electric field, which is equivalent to inserting a non-zero $\theta$-term \cite{Coleman:1978ae}. This setup corresponds to the Euclidean action\footnote{We will be interested only in the Euclidean formulation of Quantum Field Theories here.}
\be
S_{\theta}=i\frac{\theta}{2\pi}\int d^2x F\;,
\ee
where $F=\frac{\epsilon^{\mu\nu}}{2}F_{\mu\nu}=F_{01}$ is the (Euclidean) field strength. As $\theta$ moves from zero to $2\pi$, the vacuum pair of positive and negative particles can move to infinity, exactly cancelling the background electric field. When $\theta=\pi$, the background electric field corresponds to exactly half of the electric-string-flux. For large and positive scalar $M^2$ the, vacuum is twice degenerate, corresponding to the positive and negative directions of the half-electric flux, and hence breaks the charge-conjugation symmetry $\C$ spontaneously. Charged particles can be thought of as changing the electric flux by one unit, so half of the electric flux directed to the right can be absorbed by a negative charge, leaving the left-pointing half-electric-flux vacuum on the right. Such a particle is a domain-wall connecting the two vacua. However if the charge was positive, it would produce a $3/2$ electric flux on the right, which would correspond to an excited vacuum -- i.e. a string. The presence of the string is energetically penalized by the electric charge $e^2$ and the length of the string. In the limit of $e^2\rightarrow \infty$, only domain-wall excitations remain, whose statistical weight is penalized only by the length of their wordlines, and not the area they enclose. As $M^2$ is decreased, the domain walls become more common in the ensemble, mixing the two vacuua and causing $\C$ to be restored. The nature of the transition is the Ising transition \cite{Affleck:1991tj,Komargodski:2017dmc,Gattringer:2018dlw}.

But the ensemble picture, in the limit $e^2\rightarrow \infty$, looks compellingly similar to the Ising model itself, which can be thought of as the ensemble of domain walls connecting the two Ising vacua. Indeed we will see that there is a lattice gauge formulation of the gauged XY model where this identification becomes exact. The spirit of our discussion is much in the vein of lattice dualizations to worldline and worldsheets, which found recent applications for solutions of the numerical sign-problem (see \cite{Gattringer:2016kco} for a review). Moreover the formulation allows for a higher-dimensional generalization in terms of higher-form lattice gauge theories. We focus on the 2d model first for simplicity, which will render the higher-dimensional generalizations straightforward. We discuss these at the very end.

\vspace{.3cm}
\noindent{\bf The 2d gauged XY-model as an Ising model: } Let us consider a 2d lattice $\Latt$ which is made out of sites $x$, the  bonds or links $l$ and faces or plaquettes $p$. The XY model can be defined by the phases $\varphi_x\in[0,2\pi)$ living on lattice sites. It is useful to define the derivative living on the oriented link $l(x,y)$ as
\be
(d\varphi)_{l(x,y)}=\varphi_y-\varphi_x\;.
\ee 
We can write the partition function of the XY-model as
\be\label{eq:Z_XY}
\prod_{ x}\left(\int_0^{2\pi} d\varphi_{ x}\right) e^{\Jxy\sum_{l}\cos\big((d\varphi)_l\big)}\;,
\ee
where the sum in the exponent is over links of fixed orientation.
To gauge the model we introduce a link gauge field $A_l\in \mathbb R$. We further define 
\be
F_{p}=(dA)_p\equiv A_{l_1}+A_{l_2}+\dots+A_{l_i}\;.
\ee
where the links $l_1,l_2,\dots l_i$ make the boundary of the plaquette $p$. 

We take the action for the gauge fields to be
\be
S_{gauge}=\frac{\beta}{2}\sum_p (F_p+2\pi n_p)^2-i\theta \sum_p n_p\;,
\ee
where $n_p$ are integer variables on plaquettes, and where the orientation of the plaquettes is fixed in advance. The first term of the action is the famous Villain action \cite{Villain:1974ir} which has a long history and found much application in studies of abelian gauge theories and spin systems (see e.g. \cite{Peskin:1977kp,Banks:1977cc,Elitzur:1979uv,Cardy:1981qy,Svetitsky:1982gs}). Recently it also found some modern application relating to the $\theta$-terms, anomalies and formulations of generalized electromagnetic theories \cite{Gattringer:2018dlw,Sulejmanpasic:2019ytl,Gattringer:2019yof,Sulejmanpasic:2020lyq,Honda:2020txe}. 

The coupling to the XY-model is made by promoting $(d\varphi)_l\rightarrow (d\varphi)_l+A_l$ in the exponent of \eqref{eq:Z_XY}, so the partition function is now
\begin{multline}\label{eq:Z_XY_gauged}
Z=\left(\prod_{x}\int d\varphi_x\right)\left(\prod_l\int dA_l\;e^{\Jxy\cos\big((d\varphi)_l+A_l\big)}\right)\\\times\left(\prod_{p}\sum_{n_p}e^{-\frac{\beta}{2}(F_p+2\pi n_p)^2+i\theta n_p}\right)
\end{multline}

Noting that  $\sum_pF_p=0$, we can perform the Poisson resummation for each plaquette
\begin{multline}\label{eq:poisson}
\sum_{n_p\in\mathbb Z}e^{-\frac{\beta}{2}\left(F_p+2\pi n_p\right)+i\theta\left(n_p+\frac{F_p}{(2\pi)}\right)}=\\=\frac{1}{\sqrt{2\pi\beta}}\sum_{m_p\in \mathbb Z}e^{-\frac{\left(m_p-\frac{\theta}{2\pi}\right)^2}{2\beta}+iF_p m_p}\;.
\end{multline}
The RHS above is nothing but the Fourier expansion of the LHS, given that the LHS is periodic in $F_p\rightarrow F_p+2\pi$. Upon summing over all plaquettes, it is not difficult to show that $\sum_pF_pm_p=\sum_l A_l (\delta m)_l$, where 
\be
(\delta m)_l=m_{p_1}-m_{p_2}\;,
\ee
with $p_1,p_2$ being the plaquettes which share the link $l$ (the sign indicates that the plaquettes sharing the same link have opposite orientations). 

If we ignore the coupling to the XY model, integrating over $A_l$ will impose a constraint that $m_p=m$ is constant for all plaquettes. 

On the other hand we have that
\be\label{eq:XY_fourier}
e^{\Jxy \cos((d\varphi)_l+A_l)}=\sum_{k_l\in\mathbb Z}I_{k_l}(\Jxy)e^{i(d\varphi_l)k_l+i A_l k_l}\;,
\ee
which is just a Fourier expansion of the LHS. $I_k(J)$ is the modified Bessel function. Upon doing this for every link, the first term in the exponent can be ``partially integrated'', i.e.
\be
\sum_{l}(d\varphi)_lk_l=-\sum_x \varphi_x (\delta k)_x\;,
\ee
where
\be
(\delta k)_x=k_{l_1}+k_{l_2}+k_{l_3}+\dots + k_{l_i}\;,
\ee
with $l_1,l_2,\dots, l_i$ being the links oriented away from the vertex $x$. Integrating over $\varphi_x$, we have that $(\delta k)_x=0\;, \forall x\in \Latt$. This is nothing but the current conservation law, demanding that the net current $k_l$ flowing out/in of $x$ is zero. 

The partition function is now made out of closed loops of current $k_l$. By integrating over $A_l$, we further impose the constraint
\be
k_l=(\delta m)_l\;.
\ee
Note that $(\delta k)_x$ is automatically satisfied given the above constraint because $\delta^2=0$. The partition function is
\begin{multline}\label{eq:gauged_XY_dual}
Z=\left(\frac{1}{ 2\pi\beta}\right)^{P/2}\sum_{\{m\}}\left(\prod_{l}I_{(\delta m)_l}(\Jxy)\right)\\\times\left(\prod_pe^{-\frac{e^2}{2}\left(m_p-\frac{\theta}{2\pi}\right)^2}\right)
\end{multline}
where we have labeled $e^2=\frac{1}{\beta}$, $\sum_{\{m\}}$ indicates the sum over all plaquette variables $m_p$, and $P$ is the total number of plaquettes on the lattice.

Now consider the limit of $e^2\rightarrow \infty$, and $\theta=\pi$. The exponent in the 2nd line above suppresses all configurations for which $m_p$ is not equal to $0$ or $1$. Therefore, up to exponentially small corrections in $e^2$, the only allowed plaquette variables are $m_p=0,1$. These will play the role of Ising spins. Let us label $\sigma_p=2m_p-1$. Moreover, note that since $I_n(x)=I_{-n}$, the dependence on $(\delta m)_l=\frac{(\delta \sigma)_l}{2}=\frac{(\sigma_{p_1}-\sigma_{p_2})}{2}$, where $p_1$ and $p_2$ are plaquettes which share a common link $l$. Further, since $\sigma_p$ only take values $\pm 1$, we can write 
\be
I_{(\sigma_{p_1}-\sigma_{p_2})/2}(\Jxy)=\sqrt{I_0(\Jxy)I_1(\Jxy)}e^{-\frac{\sigma_{p_1}\sigma_{p_2}}{2}\log\left({\frac{I_1(\Jxy)}{I_0(\Jxy)}}\right)}\;.
\ee 
Since the plaquettes $p$ are dual to the dual lattice sites $\tilde x$, the idenity above reveals that the model in question is really the Ising model on the dual lattice, with the coupling 
\be\label{eq:JI_JXY_duality}
\JI=-\frac{\log\left(\frac{I_1(\Jxy)}{I_0(\Jxy)}\right)}{2}
\ee
Moreover one can see that $h=\frac{\pi-\theta}{2\pi\beta}$ plays the role of the magnetic field. To get the Ising model at finite $h$, one must take the double scaling limit $\theta\rightarrow \pi, \beta\rightarrow 0$ such that $h$ is finite. 

Several comments are in order
\begin{itemize}
\item If in \eqref{eq:JI_JXY_duality} we take $\Jxy>0$ then $\JI>0$, so the model maps the ferromagnetic gauged XY-model to the ferromagnetic Ising model. If $\Jxy<0$ we can shift $A_l\rightarrow A_l+\pi$ to transform $\Jxy\rightarrow -\Jxy$. Now if the original lattice $\Latt$ consists of only plaquettes which have an even number of links in their boundary (e.g. a square or a honeycomb lattice), then the shift can be absorbed by the shift of the integers $n_p$ in \eqref{eq:Z_XY_gauged}. If on the other hand hand there exist plaquettes which have an odd number of links in their boundary, it is not difficult to see that the resulting ferromagnetic Ising model partition function contains a term $e^{i\frac{\sigma_p\pi}{2}}$, which can be interpreted as the imaginary magnetic field $h=i\frac{\pi}{2}$.

\item What about antiferromagnetic Ising model on a frustrated lattice? Does there exist a $U(1)$ gauge-theory, whose dual lattice is frustrated (e.g. a honeycomb lattice), which is dual to an antiferromagnetic Ising model? For real $\Jxy$ of the XY model, the answer is no. However one can always come up with a complex value of $\Jxy$ in \eqref{eq:JI_JXY_duality} which would produce a negative value of $\JI$, so that the analytical continuation of the $U(1)$ gauge theory to complex $\Jxy$ corresponds to an antiferromagnetic Ising model.

\item While we have assumed that the coupling $\Jxy$ is the same for all links, we could make them different. The relationship \eqref{eq:Z_XY} would than be valid link-wise.

\item If we did not take the limit $e^2\rightarrow \infty$, the $XY$ model is still dual to a kind of generalized Ising model, with the spin $\sigma_{\tilde x}=2m_{\tilde x}-1$ being odd integers. The action is easily obtained from \eqref{eq:gauged_XY_dual}.

\item There is nothing particularly special about the form for the XY model \eqref{eq:Z_XY}. Indeed we could have taken the action to be an arbitrary periodic function of $(d\varphi)_l$, i.e. $S=\sum_l f((d\phi)_l)$, where $f(x+2\pi)=f(x)$. Then the Bessel functions $I_k$ should be replaced by the Fourier modes $f_k=\int_{0}^{2\pi}\frac{dx}{2\pi}\;e^{-f(x)}e^{-i k x}$. If we further demand $f(x)=f(-x)$, then $f_{k}=f_{-k}$, with all $f_k$ real. Then Ising coupling would still be given by \eqref{eq:JI_JXY_duality} with the replacement of the Bessel functions $I_0,I_1$ with $f_0,f_1$\footnote{We do not need to impose $f(x)=f(-x)$, but then the dual Ising model will in general have a nonuniform imaginary magnetic field. Further imposing $f(x)=f(-x)$ does not guarantee the positivity of the Fourier mode $f_1$, so in general $\Jxy$ may have an imaginary $\pi$-part. Note that to have a particle interpretation at finite $e^2$, we also want to demand that all Fourier modes of $f_k$ of $e^{-f(x)}$ are positive. It is not clear to us what is the most general class of $f(x)$ satisfying this (see \cite{tuck2006positivity} however).\label{fn:1}}. 

\item Our choice of gauge action is also not unique, and we could have instead chosen a gauge action $S=\sum_p f(F_p+2\pi n_p)$, so long as now $f(x)$ is \emph{not} periodic\footnote{Periodic $f(x)$ would give a partition function which is identically zero for $\theta\notin 2\pi \mathbb Z$.}. Then we can Poisson resum
\begin{multline}\label{eq:gen_poisson}
\;\;\;\sum_{n_p}e^{-f(F_p+2\pi n_p)+i\theta (n_p+\frac{F_p}{2\pi})}=\\=\sum_{m_p}{a(m_p+\theta/(2\pi))}e^{-iF_pm_p}\;,
\end{multline}
where $a(k)=\int \frac{dx}{2\pi}\; e^{ikx}e^{-f(x)}$ is the Fourer transform\footnote{Similar issues of positivity and reality of Fourier modes arise here as well (see footnote \ref{fn:1}). Note that one can also consider the Wilson action with the L\"uscher $\theta$-term \cite{Luscher:1981zq,Berg:1981er} in the form \eqref{eq:gen_poisson}, with the choice $e^{-f(x)}=e^{\beta\cos(x)}\Theta(x)$, where $\Theta(x)$ is unity for $|x|<\pi$ and zero otherwise (see discussion in \cite{Sulejmanpasic:2019ytl}). However for strong enough coupling some Fourier modes for nonzero $\theta$ will become negative, which is an artifact the Villain form avoids.} of $e^{-f(x)}$. 
\end{itemize}

Could we imagine a generalized $XY$ model described above, with purely real $f(x)$ which corresponds to an antiferromagnetic Ising model? For that to happen we must have that the 1st Fourier mode of $e^{-f(x)}$ is larger than the 0th mode, so that the logarithm in \eqref{eq:JI_JXY_duality} is negative i.e.
\be
\int_{-\pi}^\pi dx \;e^{-f(x)}\cos(x)>\int_{-\pi}^\pi dx\; e^{-f(x)}\;.
\ee
However the above can never be satisfied for real $f(x)$, and so we conclude that the antiferromagnetic Ising model on a frustrated lattice cannot be obtained from a gauged generalized XY model with real couplings.

\vspace{.3cm}
\noindent{\bf The finite coupling: from Ising to Berezinskii-Kosterlitz-Thouless (BKT) transition}

As we saw when the gauge coupling tends to infinity, $e^2\rightarrow \infty$ the ferromagnetic $XY$ model is an Ising model. Let us focus on the square lattice for concreteness, whose dual lattice is also square. It is well known that the Ising model on the square lattice has a transition at coupling \cite{Kramers:1941kn}
\be
\JI^c=\frac{\log(1+\sqrt{2})}{2}\Rightarrow {\Jxy^c}(e^2=\infty)\approx 0.9117\;,
\ee
where the XY coupling at $e^2=\infty$ was obtained with the use of \eqref{eq:JI_JXY_duality}. On the other hand we know that if $e^2\rightarrow 0$, the gauge fluctuations are completely suppressed, and we can set\footnote{Actually the zero coupling condition forces $(dA)_p=0$, but there can still be a residual nonzero holonomy in case of space-time which has incontractible loops (i.e. a nontrivial 1st cohomology group) e.g. a torus. In this case the holonomies label superselection sectors of the flat-connection gauged XY model. These sectors are equivalent to an XY model with twisted boundary conditions.} $A_l=0$, reducing the model to an ordinary XY model, which has a BKT transition at 
\be
\Jxy^c(e^2=0)=1.1194\;.
\ee
\begin{figure}[t] 
   \centering
   \includegraphics[width=0.5\textwidth]{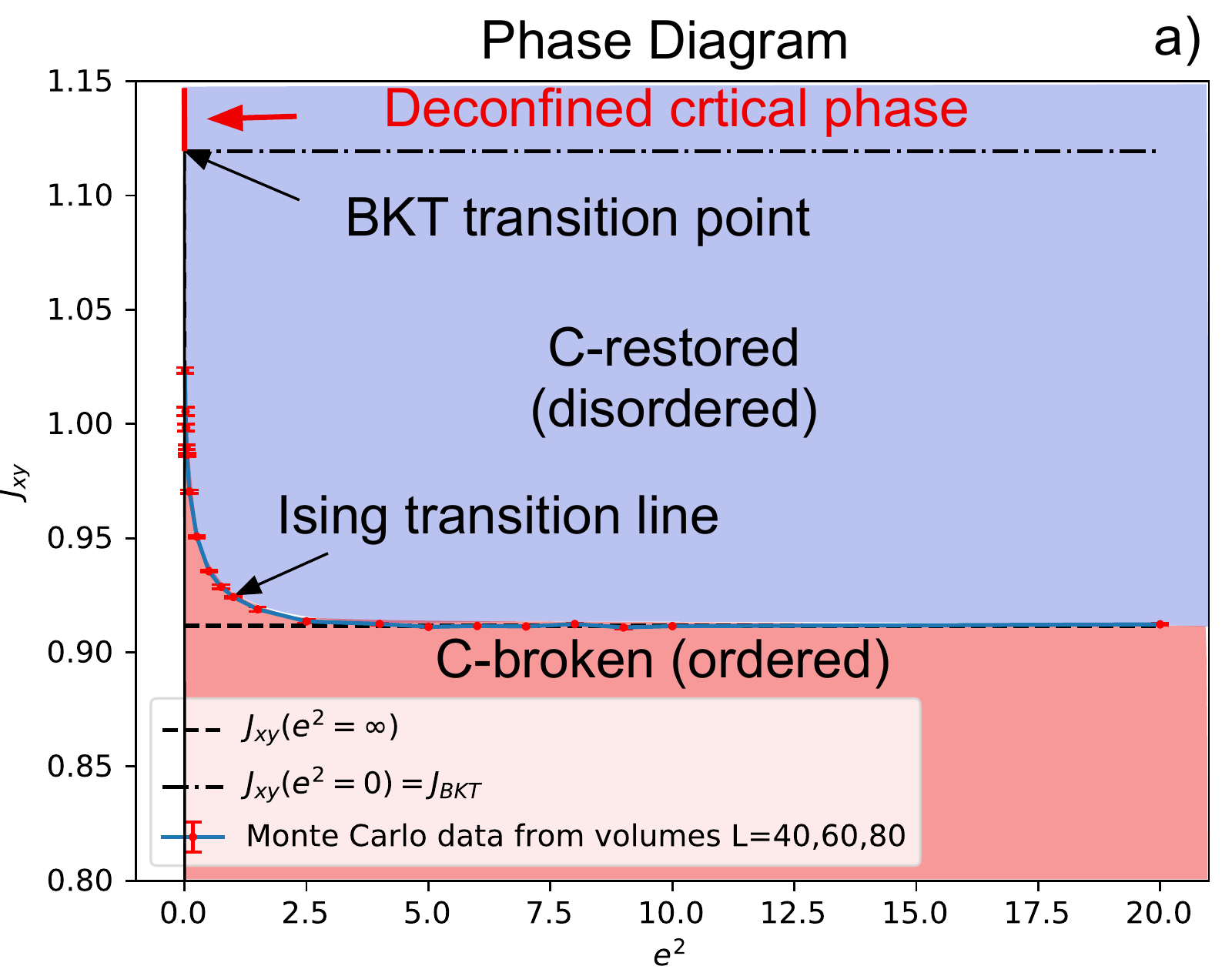}
   \includegraphics[width=0.5\textwidth]{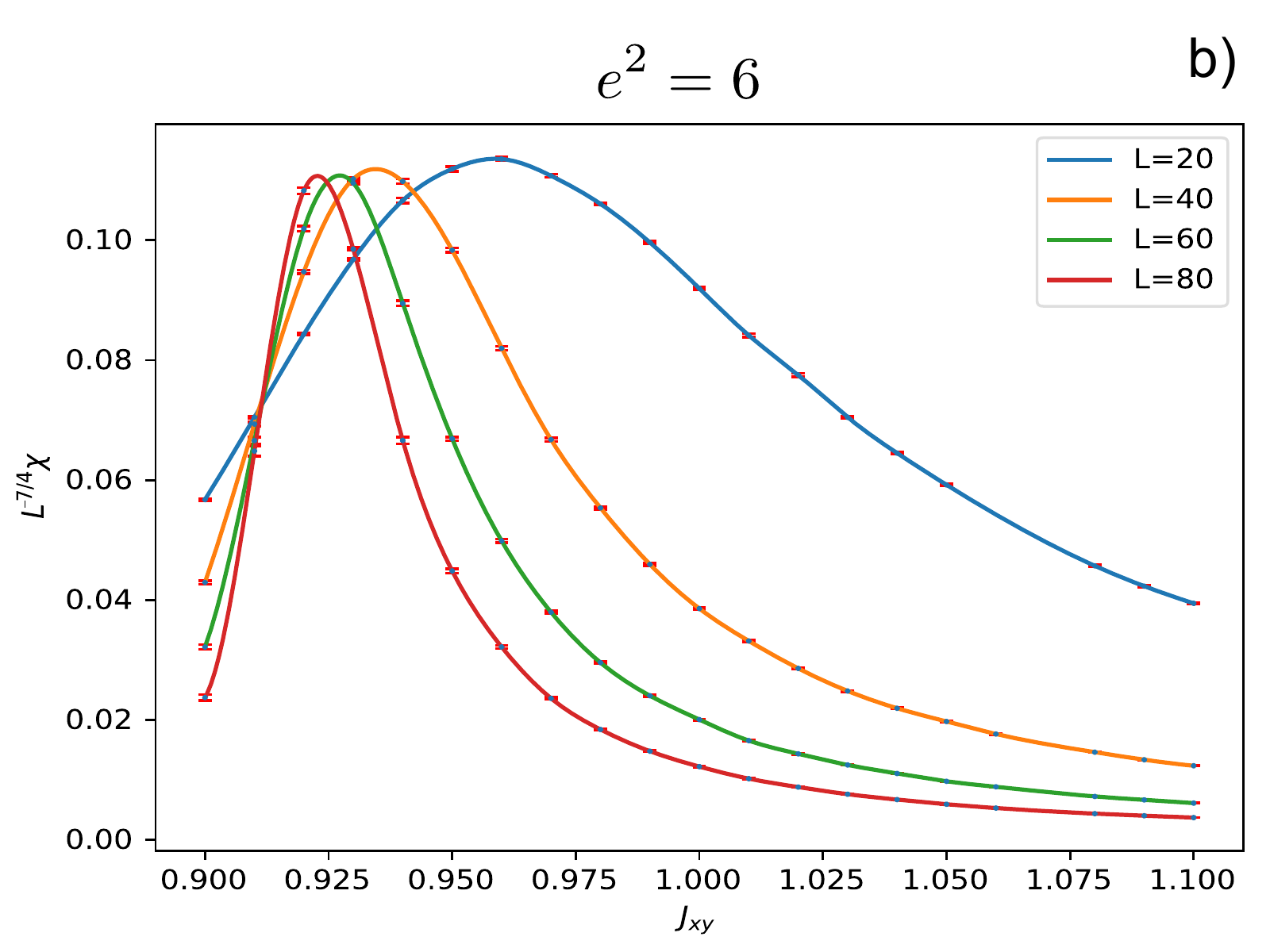} 
   \caption{a) The phase diagram of the gauged XY model. The diagram shows the transition line in the $\Jxy$ v.s. $e^2$ plane. The critical points were determined from the intersections of the rescaled susceptibility $L^{-\frac{\gamma}{\nu}}\chi_t(\Jxy)$ obtained by Monte Carlo simulations of the model \eqref{eq:gauged_XY_dual} on the square lattices with sizes $L=40,60,80$. b) An example of the rescaled susceptibility for $e^2=6$ (including $L=20$), which clearly intersect very close to a single point -- the Ising transition point. }
   \label{fig:Jxy_vs_beta}
\end{figure}

The two transitions are quite close together, differing by only $\sim 20\%$. Of course the universality class of the transitions is different. Nevertheless in both cases the transition can be thought of as the proliferation of loops. In the Ising limit, the proliferation is of the domain-wall lines, while for $e^2=0$ (i.e. $\beta=\infty$), we see that in \eqref{eq:gauged_XY_dual} the proliferation is in terms of interface-lines between different values of $m_p$-variables, which are no longer constrained to be $m_p=0,1$. Both of these proliferations are controlled by the ratio of the Bessel function $I_{(\delta m)_l}(\Jxy)/I_0(\Jxy)$, which tends to suppress the jumps in $m_p$ for smaller values of $\Jxy$, and lets them proliferate for large $\Jxy$. For intermediate $0<e^2<\infty$, the typical area of a loop bounding the region of constant $m_p\ne0$ and $m_p\ne 1$ are exponentially suppressed with $e^2$, and such domains will tend to renormalize the Ising transition, but the effect must be exponentially suppressed in large coupling $e^{-e^2(\dots)}$ (see eq. \eqref{eq:gauged_XY_dual}). On the other hand let us consider the limit $e^2\rightarrow 0$ of the XY-model in the gapped phase near the BKT transition, i.e. $\Jxy\lesssim 1.1194$. If we then change $e^2$ to be nonzero, $1/e$ will dictate the typical length-scale of gauge fluctuations in lattice units, and so it cannot induce a phase transition  until $e$ is of the order of the $XY$ mass-gap, which is exponentially small for the coupling close enough to $\Jxy=1.1194$. Hence we expect the phase transition line in the graph of $\Jxy$ v.s. $e^2$ to be slowly changing as $e^2$ is lowered from infinity, keeping close to the $\Jxy^c(e^2=\infty)=0.9117$ line, and then sharply shooting up when  $e^2$ is order unity to the value $\Jxy^c(e^2=0)=1.1194$ at $e^2=0$. 

To check this we performed a Monte Carlo simulation of the system at various values of $e^2$ on a square lattice and for the linear system sizes $L=20, 40,60$ and $80$. We define the topological susceptibility as
\be
\chi_t=\frac{1}{L^2}\frac{\partial^2 \log(Z)}{\partial\theta^2}+\frac{e^2}{(2\pi)^2}\;.
\ee
The shift by the constant above is to match the definition of the magnetic susceptibility in the Ising limit\footnote{Since the susceptibility diverges at the transition point, the constant shift affects the finite volume corrections only. This particular shift makes these correction small.}. At finite volume we expect
\be
\chi_t=L^{\frac{\gamma}{\nu}}F(tL^{\frac{1}{\nu}})
\ee
where for 2d Ising $\nu=1$ and $\gamma=7/4$ are the standard critical exponents, $t$ is the parameter driving the transition, and $F$ is the universal function. So if we plot $L^{-\frac{\gamma}{\nu}}\chi_t$ against a parameter driving the transition, we expect that, at the phase transition point $t=0$ the curves will cross. Indeed, plotting $L^{-\frac{\gamma}{\nu}}\chi_t$ against $\Jxy$ shows that all the curves intersect pretty closely at a single point, as can be seen in Fig.~\ref{fig:Jxy_vs_beta}b) where simulations for $e^2=6$, are shown for the four volumes. We repeated the simulations for values of $e^2$ ranging from $e^2=0.01$ up to $20$, to produce a phase diagram as indicated by red datapoints in Fig.~\ref{fig:Jxy_vs_beta}a). Note that we excluded the data for $L=20$ to minimize power corrections to the scaling. In addition the Ising scaling, discussed above does not set in at $L=20$ for the smallest values of $e^2$. This is expected as the dominant fixed point for small enough volumes should be of the BKT nature.

\vspace{.3cm}
\noindent{\bf Generalizations to higher dimensions: } Generalization to higher dimensional cases is now straightforward. First we define the lattice $\Latt$ in terms of $p$-cells $c_p$. A $0$-cell is a vertex. We then connect vertices with $1$-cells (links), and $1$-cells with $2$-cells (plaquettes), etc. In $D$-dimensions we define a $(D-1)$-form gauge field $U(1)$, which will naturally live on $(D-1)$-cells, which we label as $B_{c_{D-1}}$. This is the generalization of $A_l$ for the spacetime dimension $D=2$. In addition we introduce $(D-2)$-form gauge field $A_{c_{D-2}}$,  living on $c_{D-2}$. Similar to before we define the derivatives $d$ and $\delta$ which map a $p$-form field to a $p+1$ and $p-1$ form field respectively (see e.g. appendix of \cite{Sulejmanpasic:2019ytl} for details). We define the prototypical action
\begin{multline}
\sum_{c_{D}}\frac{\beta}{2}\left[(dB)_{c_D}+2\pi n_{c_D}\right]^2+i\theta n_{c_D}\\-J\sum_{c_{D-1}}\cos[(dA)_{c_{D-1}}\!+\!B_{c_{D-1}}]\;.
\end{multline}
The action is just the generalization of the exponent in \eqref{eq:Z_XY}. Note that the $\theta$-angle has a similar interpretation as before: a $D-1$-form $U(1)$ gauge field $B$ has a natural topological charge in the continuum given by $\frac{1}{2\pi}\int dB$. An example of such a gauge field is the nonabelian Chern-Simons $3$-form in 4 spacetime dimensions. Similar reasoning as before leads to the dual partition function 
\begin{multline}\label{eq:gauged_higher_dual}
Z=\left(\frac{1}{ 2\pi\beta}\right)^{C(D)/2}\sum_{\{m\}}\left(\prod_{c_{D-1}}I_{(\delta m)_{c_{D-1}}}(J)\right)\\\times\left(\prod_{c_D}e^{-\frac{e^2}{2}\left(m_{c_D}-\frac{\theta}{2\pi}\right)^2}\right)\;,
\end{multline}
where again $e^2=\frac{1}{\beta}$, and $C(D)$ is the number of $D$-cells on the lattice. Now we identify $c_D$ with the site of a dual lattice $\tilde x$, and define $\sigma_{\tilde x}=2m_{\tilde x}-1$ to be the spin variable. Then at $\theta=\pi$, in the limit $e^2\rightarrow 0$ only $\sigma_=\pm 1$ survive, and the model reduces to the $D$-dimensional Ising model with the coupling given by \eqref{eq:JI_JXY_duality}, with $J_{xy}$ replaced by $J$. 

Let us briefly discuss the phase diagram as a function of $e^2$. At $e^2\rightarrow \infty$, we have that the model undergoes a phase transition at some value of $J_c$, which corresponds to the Ising transition via the duality relation. Just like before, as $e^2$ is reduced, the phase transition is expected to raise to slightly larger values of $J_c$, similar to Fig.~\ref{fig:Jxy_vs_beta}a. However in the limit $e^2\rightarrow 0$, the model in question is the $(D-2)$-form lattice gauge theory with the standard Wilson action. For $D=3$, it is just the usual lattice gauge theory, which is well known to always be in the gapped phase \cite{Polyakov:1976fu,Polyakov:1987ez}, because the theory always has monopoles. However these are expected to be suppressed exponentially with $J$, and so for very large values of $J$, the mass gap $M$ will be exponentially small. The introduction of nonzero $e^2$, where $e$ again has an interpretation as the length of the $B$-field fluctuations, will therefore be able to induce a transition only when $e$ is of the order of $M$, which is tiny. So the qualitative picture is very similar to Fig.~\ref{fig:Jxy_vs_beta}a), except that the phase-transition boundary shoots up to infinity for $e^2\rightarrow 0$.

\vspace{.2cm}

\noindent{\bf Acknowledgments: } I would like to thank Bernard Piette for helping me understand how to use the Condor computer cluster at the Department of Mathematical Sciences, Durham University. I would also like to thank Christof Gattringer for useful comments on the manuscript, and, together with Daniel G\"oschl and Nabil Iqbal for input with regards to the Monte Carlo error analysis. This work is supported by the Royal Society of London University Research Fellowship.

\bibliographystyle{utphys}
\bibliography{bibliography}


\end{document}